# A memory mechanism based on two dimensional code of neurosome pattern

Shengyong Xu and Jingjing Xu

Key Laboratory for the Physics & Chemistry of Nanodevices, and Department of Electronics, School of Electronics Engineering and Computer Science, Peking University, Beijing, 100871, P. R. China

Emails: xusy@pku.edu.cn (SYX), xujj@pku.edu.cn (JJX)

"*How Are Memories Stored and Retrieved?*" This question remains puzzling to date [1-12] . Besides human, apes, elephants, dolphins and octopuses are well-known smart animals with limited intelligence and memory capability. Some eagles know how to throw a thigh-bone from air onto a rock so to break the bone and take its marrow. Even an ant remembers its home and is able to collaborate with partners in hunting foods and building a nest. The memory mechanism of a human brain and animal brains most likely follows the same basic principle, thus it might be simple.

Here we present a model for the material foundation of memory. The model is based on facts that we have known for neurons, synapse junctions and neurotransmitters in a brain. When a neuron cell is excited, its neurosome, dendrites and axon(s) are all consequently excited, where measurable transmembrane ionic ($Na^+$, $K^+$, $Ca^{2+}$) currents are generated at its protein channels and neurotransmitters are released at its synapse junctions [13,14] . In general, based on measurable properties synapse junctions are cataloged into electrical synapses (ES), chemical synapses (CS) and mixed (i.e., partial electrical and partial chemical) synapses (MS). Under certain conditions, an ES (or MS) can turn into a CS, and vice versa [15-18] . Hereinafter, when two neurosomes are connected with an ES, they are referred as "strongly connected", and excitation of either one neurosome directly leads to the excitation of the other without delay. If two neurosomes are connected with a CS or MS, their connection is referred as "loosely connected". The transmission of excitation from one neurosome to the other through a CS could be nonreversible, or even blocked when receiving inhibitory transmitters from a third neurosome [19] . The connection strength of a MS is considered in between that of an ES and a CS.

In a human brain, its cortex layer has a large surface area, and neurosomes in the cortex are structured into six or more two dimensional (2D) layers [20,21] . The visual information of a picture, for instance, is recorded and turned into electrical neural pulses via light sensitive cell-sensors on retina. The electrical pulses are then transmitted through array of axons, finally received by a 2D array of neurosomes in the cortex of the brain. Similarly, the information of a sound, a smell, a touch of the body, etc., also appears as a pattern of excited neurosomes at certain regions of the cortex [12,22].

We believe that the basic information data of memory in a brain are stored in the form of 2D codes. Each 2D code is a pattern of 2D network, which consists of a number of strongly connected neurosomes. When any one neurosome of this strongly connected 2D network is excited, other neurosomes of the whole network will be excited one by one in a very short time (at the order of 0.1 ms), nearly simultaneously. As a result, the geometric 2D pattern of the exciting neurosomes makes up a single 2D code as the information carrier for memory and other brain functions. Except for some instincts, where patterns of exciting neurons as well as the body reactions are built-in at birth, to recognize the meaning of 2D codes in memory needs learning processes.

We define a sensory layer of neurosomes in the cortex as the layer receiving directly the electrical pulses from sensors of eyes, nose, ears, skin, etc. Since the sensory layer needs to reset again and again to keep fresh for receiving continuous electrical pulses, it seems not capable of serving the same time as data analysis layer or the memory layer. A feasible way is to project the pattern of excited neurosomes, i.e., 2D codes of data, onto a layer close to it for analysis and onto another layer or layers for temporary memory, via vertical dendrites and/or axons. Figure 1a schematically illustrates the layered structure of neurosomes for different functions.

Yet an electrical pulse of an excited neurosome lasts only around 1 ms [13]. This might be too short for both analysis and memory processes. To keep the 2D pattern of excited neurosome remaining in the memory layer(s), an echoing mechanism may work well. It needs two layers of neurosomes that are vertically connected with CSs or MSs. The 2D patterns of excited neurosome are copied in the first memory layer then projected to the second memory layer via vertical dendrites. Next, the patterns are projected back to the first layer. By repeating this process, the information stored in the 2D patterns of exciting neurosomes are echoed backwards and forwards between the two memory layers, forming an oscillating state for a certain period. In this way, a temporary memory is established for retrieving or analyzing functions.

Since the 2D pattern of exciting neurosomes in the sensory layer is a direct projection of the 2D array of ascending sensory nerve fibers, each neurosome in the sensory layer is expected isolated to each other so to avoid data interference. On the other hand, the memory layers need to store a large number of 2D codes of temporary memory in similar locations in order to let the brain be capable of thinking slowly and analyzing complex information. Thus ideally one neurosome in the sensory layer is connect to a number of neurosomes in the memory layers via dendrites and synapses.

To copy a 2D code of exciting neurosomes onto another layer with exactly the same pattern as that in the sensory layer, it should require a special interaction rule among synapses. One of the feasible choices is the following: when two loosely connected neurosomes are simultaneously excited, their interconnection junctions receive the electrical pulses from both sides simultaneously, then the junction is going to be greatly strengthened by turning a CS or MS into an ES with a large probability. A threshold of local electric field may exist at the synapse region, where the turnover from a CS (MS) to an ES occurs. For example, the threshold could be higher than the value generated by a single electrical pulse, but lower than twice of the value. In other cases, when only one of the two loosely connected neurosomes is excited and sends electrical pulses to their synapse junction, the other neurosome just experiences a secondary excitation, but their synapse junction remains unchanged.

As shown in Figure 2, by this approach, only when two loosely connected neurosomes simultaneously receive electrical pulses from the sensory layer and simultaneously send electrical pulses to each other, their interconnections turn from a CS or MS junction into an ES junction. As a result the 2D code in the sensory layer is copied accurately into the memory layers with the same geometric pattern.

Furthermore, statistically a single projection action from the sensory layer to the memory layers may only lead to a poor copy accuracy of the 2D code. The echoing process between two memory layers may also cause error and lose part of the information. This is consistent with the typical nature of temporary memory. In a learning process, when the same 2D codes are repeatedly projected to the memory layers, it will result in better copies of the codes and longer lasting memory strength. If the learning process is repeated enough times within a certain period, the 2D codes can be finally consolidated, thus turning temporary memory into long-term memory. The threshold of learning process for a long-term memory may depend on the complexity and number of the 2D codes to be memorized.

On the other hand, the strength of interconnection junctions among neurons are observed decreasing with time [5] . Therefore, temporary memory is gradually lost when one or more neurosomes in a strongly connected 2D network lose the strength in interconnections with partners, i.e., turning from an ES to a MS or CS. For long-term memory, this process is slower as the threshold for the reversal process is much higher than that for losing temporary memory. Clearly, when some neurons involved in 2D codes die or are damaged, the memorized information is then partially or totally lost [23,24] . Yet rewiring the strongly connected neurosome pattern of a 2D code is possible. New connections could set up so that the dead neurosome is bypassed or replaced in new 2D code.

To avoid interference between neighboring 2D networks of strongly connected neurosomes, it is reasonably assumed that each neurosome is only involved in one 2D code. This leads to limited storage capacity of 2D codes in a brain. A human brain consists of around 10 billion neurons. If one 2D code takes averagely 10 neurosomes, the maximum number of 2D codes that could be stored is thus about 1 billion.

An analysis layer of neurosomes is necessary for analyzing 2D codes projected from the sensory layer, as well as for retrieving 2D codes from memory layers. And, in the analysis layer the 2D codes from different cortex areas like vision, olfaction, audio, touch, emotion and events, etc., are to be analyzed and compared at the same time and same region. The phenomenon of synesthesia is a typical case. It is the foundation for learning, thinking and reasoning functions of a brain.

Retrieving memory data is supposed to perform in the analysis layer. In a retrieving process, each 2D code in the memory layers is triggered via dendrites and/or axons connected to the analysis layer. Here certain category functions of 2D codes in memory layers and related searching functions connecting the analysis layer to memory layer should be established in a brain, so that memorized 2D codes can be found quickly. The exact working functions are not clear. These functions are closely related to the physical structures for consciousness and logical thinking, which are beyond the scope of this paper and we will discuss them elsewhere.

Once a memory is searched and triggered, one or a series of 2D codes will be excited and projected to the analysis layer in a time sequence. In the analysis layer, comparison function is supposed to perform frequently. For instance, when one neurosome in this layer receives simultaneously two electrical pulses from two neurosomes of the sensory and memory layer, respectively, it may undergo an electrical field higher than a threshold and thus is triggered into an excitation state (Figure 1b), resulting a positive output. But if only receiving one electrical pulse, it is not excited or leads to a negative output. In this way the neurosome serves as an "adder". Such a function assembles the triggering mechanism of capture action of a flytrap. Once any two of six sensory hairs on the trap leaves are touched twice within 20-30 s, the trap is able to close in 100-300 ms after the second touch [25,26]. The "adder" function helps the analysis layer to make comparison. Logically, when two pieces of 2D codes retrieved from different resources are projected to the same array of neurosomes, the more overlapping points occurs, the more number of neurosomes are triggered to excite. The outcomes from the analysis layer may be stored in additional memory layers with a similar echoing mechanism for further brain functions. In history human beings developed a variety of symbols such as vocal sounds and written characters to extend the storage volume of memory. The manmade symbols play an irreplaceable role in the development of human brain, finally making it superior to animal brains.

In short, we have recognized that 2D codes, i.e., a group of strongly connected neurosomes that can be simultaneously excited, are the basic data carriers for memory in a brain. An echoing mechanism between two neighboring layers of neurosomes is assumed to establish temporary memory, and repeating processes enhance the formation of long-term memory. Creation and degradation of memory information are statistically. The maximum capacity of memory storage in a human brain is estimated to be one billion of 2D codes. By triggering one or more neurosomes in a neurosome-based 2D code, the whole strongly connected neurosome network is capable of exciting simultaneously and projecting its excitation onto an analysis layer of neurons in cortex, thus retrieving the stored memory data. The capability of comparing two 2D codes in the analysis layer is one of the major brain functions.

**Acknowledgements**

We thank Dr. J. Tang and Dr. R. J. Dai for valuable discussions. This work is financially supported by Peking University and the National Natural Science Foundation of China (NSFC Grants No. 11374016).

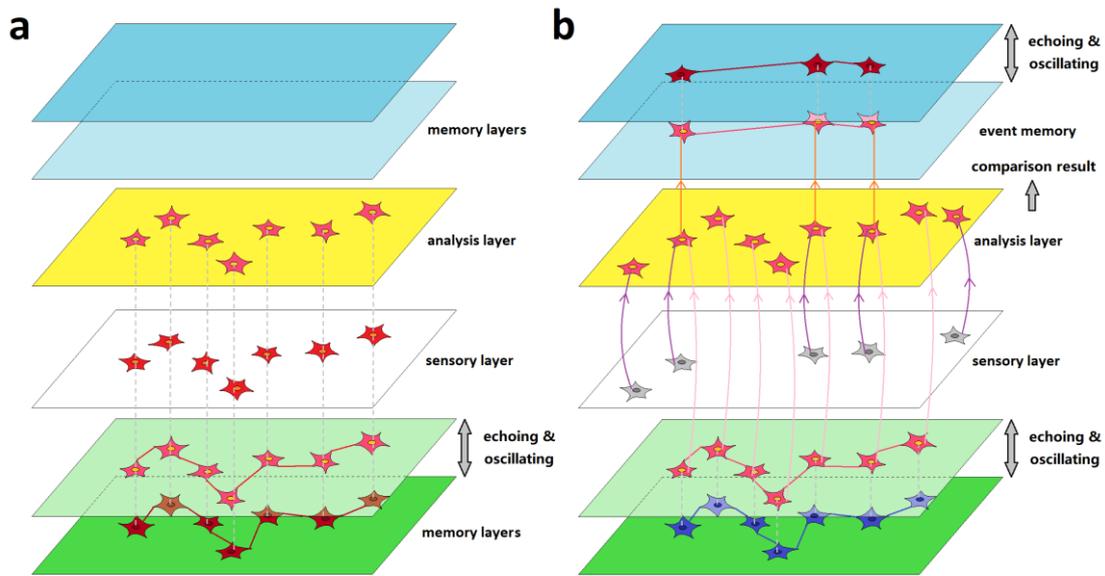

**Figure 1**. Schematic illusion of multilayered neurosome structures for creating memory and retrieving memory in a brain. (a) The information data appear as 2D patterns of excited neurosomes in the sensory layer. By projecting the 2D pattern into memory layers through dendrites, and by an echoing process, strongly connected 2D pattern of neurosomes can be created and maintained in the memory layers, following the exact 2D pattern in the sensory layer. (b) In a retrieving process, the 2D codes in memory layers are projected onto analysis layer. Comparison function can perform here to compare data stored in neurosome patterns from sensory layer (shown in grey color) with that of memory layer (shown in dark blue). The curved solid lines with arrows indicate projection of the 2D patterns onto analysis layer via dendrites or axons. The comparison results can be stored in additional memory layers next to the analysis layer. Vertical dash lines in the figure indicate the dendrite connection between layered neurosomes.

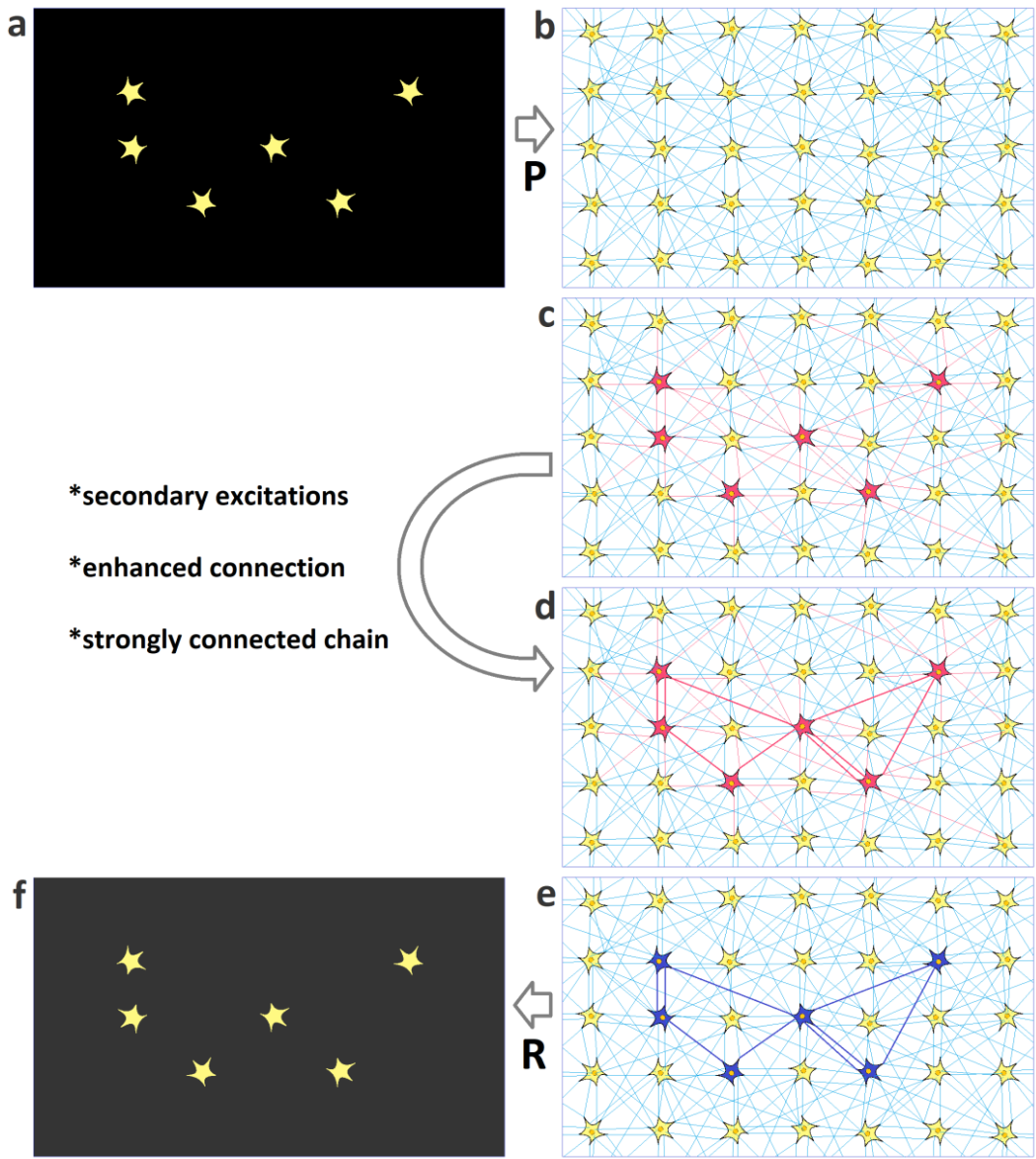

**Figure 2**. Schematic illustration for the formation and retrieval of 2D codes of temporary memory. (a) A 2D pattern of individually excited neurosomes in the sensory layer. They are direct displays of the electrical data transmitted from sensors of the body. (b) Original neurosome array in the memory layer, where neurosomes are loosely connected with CS and MS (highlighted with light blue lines). (c) When receiving projection (highlighted with "P" and an arrow) of excitation of the 2D pattern in the sensory layer through vertical dendrites, some neurosomes are excited first. (d) Formation of strong connections between two neurosomes excited simultaneously, where CS (or MS) is turned into ES. (e) After oscillation between two memory layers and repeating learning processes, a strongly connected network of neurosomes (highlighted with dark blue) is consolidated, forming a stable 2D code for temporary or long-term memory. (f) Once triggered, the 2D code stored as a strongly connected pattern of neurosomes is projected onto the analysis layer, thus retrieving the memorized data.